\documentstyle[aps,amssymb,psfig]{revtex}
\begin{document}
\title{Finite-lattice expansion for Ising models on quasiperiodic tilings}

\author{Przemyslaw Repetowicz}
\address{Department of Mathematics,Heriot-Watt University,Riccarton, Edinburgh EH14 4AS,United Kingdom}

\date{\today}
\maketitle

\begin{abstract}
Low-temperature series are calculated for the free energy, magnetisation, susceptibility
and field-derivatives of the susceptibility in the Ising model on the quasiperiodic Penrose lattice.
The series are computed to order $20$ and estimates of the critical exponents $\alpha$, $\beta$ and
$\gamma$ are obtained from Pad\'{e} approximants.
\end{abstract}


\section{Introduction}
The problem of the relevance of disorder for phase transitions in
lattice models of statistical mechanics has attracted attention for many years
and the discovery of quasicrystals \cite{Schechtman} has served to increase interest 
in the physical properties of disordered systems. A fundamental problem in this field is 
whether quasiperiodic order
is strong enough to change the critical behaviour of magnetic phase transitions.
To investigate this problem we consider in this article a classical Ising model defined on an underlying
quasiperiodic lattice.

There have been many works in this field since  the late eighties.
A heuristic criterion (Harris-Luck criterion) has been formulated \cite{Luck} which relates the 
critical behaviour to fluctuations of the number of spin couplings in a given region. The spatial scaling 
of fluctuations was described in terms of a ``wandering exponent'' $\omega$ which was 
required to exceed a threshold $\omega_c$ in order to produce a new universality class.
For the majority of quasiperiodic structures existing in reality, such as the structural
models of quasicrystalline phases  discovered so far $\omega$ 
can be calculated exactly, due to the self-similarity or inflational symmetry 
of the structure, yielding a value smaller than the threshold and suggesting the irrelevance
of disorder. However since numerous structures like the rhombic sevenfold or ninefold lattices 
\cite{RepetSeven,Danzer} exist,
 which are deprived of inflational symmetry and are therefore
 potential candidates for novel critical behaviour, 
there is still a strong motivation for dealing with quasiperiodic Ising models.

Quasiperiodic Ising models were investigated by Monte-Carlo simulations \cite{OkaNii1,Sorensen,Ledue,Oli}
which at present, seem to yield
the most precise estimates for the transition temperature and critical exponents.
Indeed, in \cite{Sorensen} computations for large periodic approximants (PA)
of the Penrose tiling (PT) \cite{Penrosetiling} were carried out 
and obtained values for the correlation length $\nu$ and the
two-spin correlation function $\eta$ exponents with two-digit precision 
($\nu = 1.02\pm 0.02$, $\eta=0.252\pm0.003$) which agreed with the square lattice values
($\nu = 1$, $\eta=0.25$). Moreover, the non-universal critical temperature $T_c$ 
has also been determined with an impressively small error $k T_c = 2.398\pm0.003$.

It is worth mentioning that a novel invaded-cluster algorithm, which modifies the temperature 
during the simulation towards the critical one, as opposed to standard Monte-Carlo algorithms with fixed temperature,
was also applied
to quasiperiodic systems \cite{Oli} to give an improved estimate of $T_c$.
The critical exponents are not available in this case, however.

Another approach is an approximate renormalisation group analysis 
\cite{Bose,AoyOda} which yields poor results, however.
 For the PT the specific heat exponent equals $\alpha=-0.1083$ versus $\alpha=0$ for the square lattice.

Quasiperiodic Ising models were also examined by graphical expansion methods \cite{AbeDotera,RepetGrimm} 
and by calculating exact partition functions for PA, obtained
from the Kac-Ward determinant \cite{RepetGrimm1}.
In the first case estimates for $T_c$ and critical exponents have not been considerably improved but
this approach demonstrated a new feature, a very slow convergence of the partition function $(Z)$ series to
its predicted asymptotic form. We also investigated the set of zeros of $Z$ 
in the complex plane (Fisher zeros), which
turned out to be much more complicated than in the square lattice case.

The Kac-Ward determinant method appeared to yield highly accurate estimates 
of the critical temperature of quasiperiodic Ising models 
(for example $k T_c = 2.397820(7)$ for the PT).
Moreover, within this framework it was possible to construct a two dimensional Ising model
with relevant fluctuations, i.e. for which $\omega > \omega_c$, which shows up another novel feature, 
namely the divergence
of high temperature series \cite{DipRepetowicz}. This example is interesting because it
points out that in some cases 
the reliability
of methods for extracting critical values from analysis of a series expansion,
like the Pad\'{e}- or differential-approximants methods, 
\cite{DombGreenI}
can be questioned.
An inspection of the Fisher zeros furnished the explanation, 
since it appeared that the moduli of some complex zeros
were smaller than the modulus of the physical singularity (real zero) and thus the complex zeros, 
rather than the real zero, were limiting the
region of convergence the series.

In this paper it is not our purpose to improve on estimating $T_c$ or $\alpha$ for 
quasiperiodic Ising models
since, due to the slow convergence of series expansions \cite{RepetGrimm1} a large, inaccessible
 number of terms
is needed to make progress in this field.
Instead, we aim at generalising the series expansion approach to the case of non-zero field quasiperiodic 
Ising or Potts models \cite{PottsModel} and provide alternative estimates of the magnetic exponents, 
$\beta$, and $\gamma$. 
Moreover, this approach allows us to investigate the problem of a disorder-driven ``softening'' 
of the first-order phase transition
in $Q$-state Potts models for $Q>4$ \cite{JankeJohnston}.

\section{The finite lattice expansion method for Ising models}
The problem consists in calculating the partition function $Z({\cal G})$ of an Ising model
on a lattice ${\cal G}$ by series expansion.
The partition function with field $B$ and coupling constant $J$ is defined in the usual way:
\begin{equation}
Z({\cal G}) \;=\; \sum_{\{\sigma_j\}} \exp\beta \{- J\sum_{\langle j,k\rangle} 
                  \Delta(\sigma_j,\sigma_k) - B\sum_{j=1}^N \Delta(\sigma_j,0) \}
\quad\mbox{where}\quad 
\Delta(\sigma_1,\sigma_2) \;=\; \left\{
                               \begin{array}{cc}
                                0 & \sigma_1 = \sigma_2 \\
                                1 & \mbox{otherwise} \\
                               \end{array}
                                \right.
\end{equation}
where the sum over spin configurations $\{\sigma_j\} = \{\sigma_1,\sigma_2,\ldots,\sigma_N\}$
consists of $N$ sums each of which runs over $\sigma_j=\{1,2\}$. 
Starting from cluster integral theory (page 42--46 and page 73 in \cite{DombGreenII}) one can formulate
a free energy (F)  expansion in terms of connected graphs 
for a wide range of models from statistical mechanics.
In particular, for the non-zero field Ising model or the $Q$-state Potts models 
(the generalisation of the former one with $Q$ values of spin at each site)
the expansion on a lattice ${\cal G}$ reads
\begin{equation}
\log{Z({\cal G})} \;=\; \sum_{r} (C_r ;{\cal G})\, k_r(w) \quad\mbox{where}\quad w = \tanh{\beta J}
\label{eq:high_temp_expans}
\end{equation}
where the sum on the right-hand side runs over connected graphs $C_r$ from ${\cal G}$.
The quantity $(C_r ;{\cal G})$ denotes the embedding number of $C_r$ in ${\cal G}$,
counting the number of ways $C_r$ can be embedded in ${\cal G}$. Finally, the weight functions
$k_r(w)$ depend only on $C_r$ not on $G$.
Making use of the independence of weights $k_r(w)$ from the lattice we can write equation
(\ref{eq:high_temp_expans}) substituting each connected graph $C_r$ for ${\cal G}$, solve the system of equations
for the weights and plug in the results to equation (\ref{eq:high_temp_expans}).
We obtain
\begin{equation}
\log{Z({\cal G})} \;=\; \sum_{g_r} a_r \log{Z(g_r)}
\label{eq:high_temp_expans1}
\end{equation}
where $a_r = \sum_p (g_p ; G) b_{r,p}$ and $b_{r,p}$ is inverse to the matrix of embedding numbers,
i.e. $b_{r,p} = (g_r ; g_p) ^{-1}$.
The sum on the right-hand side in (\ref{eq:high_temp_expans1}) runs over $g_r$ from a subset of all connected
graphs. It turns out \cite{HijmBoer} that the graph $g_r$ can furnish a non-vanishing contribution, i.e. $a_r\ne 0$,
if and only if it is an overlap of the embeddings of two other graphs having non-vanishing contributions.
The construction of graphs therefore runs as follows; we start from several ``fairly large'' graphs
and construct all possible overlaps of their embeddings in the lattice in a recursive way.
This limits the number of contributing graphs considerably, 
when compared to expansion (\ref{eq:high_temp_expans}), but, except for regular lattices like the square
or honeycomb lattice, still leaves the problem of determining $g_r$ and the contributions $a_r$ open.
Indeed, for the square lattice where $g_r$ are rectangles, $a_r$ can be explicitly expressed via
the ratio of the graph side lengths \cite{Enting} and the order to which the expansion is correct 
is in direct connection with the perimeter length of the largest graphs under consideration.
For the quasiperiodic lattices which we wish to investigate the problem is not so simple.
In what follows we focus on the PT \cite{Penrosetiling} and present the details
of the expansion method for it in the next section. 

\section{Calculation of series expansion for the Penrose tiling}
\label{sec:Calculation}
The PT is an aperiodic tiling of a plane by two kinds of rhombi of unit length side 
with angles $2 \pi/5$ and $4 \pi/5$ respectively. A discussion of the methods of generation and 
geometrical properties of this tiling can be found in \cite{RepetGrimm}, here we only mention
a particularly useful feature, namely that embedding numbers of finite patches from this tiling can be calculated 
exactly and take the form $n + m\tau$ where $\tau = (\sqrt{2}+1 )/2$ is the golden number 
and $n,m$ are rational numbers. The calculation of the series expansion consists therefore
of the  following steps:
\begin{enumerate}
\item Choose an initial set of ``fairly large'' graphs 
      which are expected to be
      large enough that every connected subgraph of the 
      underlying tiling with perimeter length not larger than a given threshold $2 L$
      can be embedded in one of them.
      While on the square lattice this condition is satisfied by all possible rectangles 
      with perimeter length $2 L$ on the PT the things are worse due to the 
      lack of periodicity of the tiling. Moreover, as opposed to the square lattice 
      graphs in the PT can have 
      different ``boundary line fillings'', i.e. there are different graphs having the same 
      boundary line \cite{RepetGrimm}.
      Knowing that the PT contains eight different vertex types, i.e. different
      site environments related to the nearest neighbours, we cut out appropriately
      large patches around each vertex type, obtaining eight patches, and found 
      all possible ``boundary line fillings''.
      There is still a lot of ambiguity in this procedure since a patch in not uniquely 
      determined by the vertex type of its central site. It would be more correct to take
      all possible higher order vertex types \cite{BaaGriReJo}, i.e. $m$-order vertex types 
      related to neighbours
      located not further than $m$ edges lengths from the site, but since their number grows
      quite rapidly with the order, the initial set of graphs would be too numerous and
      the generation of overlaps (see next item) too time consuming.   
\item Generate all possible overlaps of embeddings of the initial graphs in the tiling.

      It is difficult to estimate how the time of the generation depends on the number
      of initial graphs. Let us say a couple of words about this, however.
      We group graphs into generations so that the initial set of graphs constitutes the zeroth
      generation and the nth generation consists of overlaps of graphs from the (n-1)st 
      and zeroth generations. Since the time for creating the nth generation depends
      on the product of numbers of graphs from generation zero, $\#g(0)$ and generation $(n-1)$, $\#g(n-1)$,
      starting from a too numerous 
      zeroth generation should be avoided. The total number of overlaps grows rather slowly 
      with $\#g(0)$ for large $\#g(0)$ and most of the computing time will be devoted
      to checking and rejecting graphs which occurred before. On the other hand if we took
      too few initial patches, the covering of the lattice with them would be incomplete, 
      there would be plenty of ``holes'' not covered by any of the patches, and thus the 
      series expansion would be error laden. The rule of thumb is to take $g(0)$ not larger than
      twenty and choose the patches in such a way that their interiors differ as much as possible.

      Again, on the square lattice it's immediately clear that the overlaps are rectangles,
      because every rectangle can be constructed as an overlap of two other rectangles,
      whereas on the PT the shapes of graphs and their quantity depends on the initial set. 
      To make the things worse we are not even sure that we obtain star graphs 
      (page 1--16 in \cite{DombGreenII}), i.e. graphs without articulation points, 
      because the initial graphs are not necessarily convex.
      Connected graphs consisting of multiple components will cause
      some difficulties by the calculation of partition functions by the transfer matrix method
      (see following items).
\item Calculate the contribution $a_r$ of graph $g_r$ in (\ref{eq:high_temp_expans1})
      in the following recursive way 
      \begin{equation}
      a_r \;=\; ( g_r ; G ) - \sum_{r \in p} a_p ( g_r ; g_p )
      \label{eq:graphcontrib}
      \end{equation}
      where the sum on the right-hand side runs over all graphs $g_p$ in which $g_r$ can be embedded.
\item Calculate logarithms of partition functions $\log{Z(g_r)}$ by the transfer-matrix method.
\end{enumerate}      
\centerline{{\bf The transfer-matrix method}}
      Here we have to distinguish two cases, namely the case when the graph has no articulation points (star graph)
      and the contrary (multicomponent graph). The latter is undoubtedly more complicated
      but fortunately it turns out that it takes place only in a minority of the graphs under consideration.
      Let us firstly discuss the case of a star graph.
      We can define a perimeter of the graph, i.e. a line consisting of edges each of which belongs only to one rhombi.
      The sum over spin configurations can be performed by moving a boundary line across the graph. 
      At each stage the boundary line goes
      through a number, say $k$, sites. For the $Q$-state Potts model we have $Q^k$ different spin configurations
      on the boundary line. 
      Now we define a $Q^k$ dimensional vector $Z({\bf \sigma})$ consisting of partition functions calculated for the patch
      composed of sites from the boundary line, with a given spin configuration 
      ${\bf \sigma}=\{\sigma_1,\sigma_2,\ldots,\sigma_N\}$ assigned to them, 
      and sites already traversed by the boundary line. 
      The initial values of $Z({\bf \sigma})$ are given by:
      \begin{equation}
      Z({\bf \sigma}) \;=\; \tilde{x}^a (1-\tilde{y})^b
      \quad\mbox{where}\quad   \tilde{x} \,=\, \exp(-\beta J),\quad \tilde{y} \,=\, 1 - \exp(-\beta B/2)
      \end{equation} 
      and
      \begin{equation}
      a=\sum_{p=1}^{N-1} \Delta(\sigma_p,\sigma_{p+1}), \quad b=2 \sum_{p=1}^{N} \Delta(\sigma_p,0)
      \end{equation} 

\begin{figure}
      \centerline{\psfig{figure=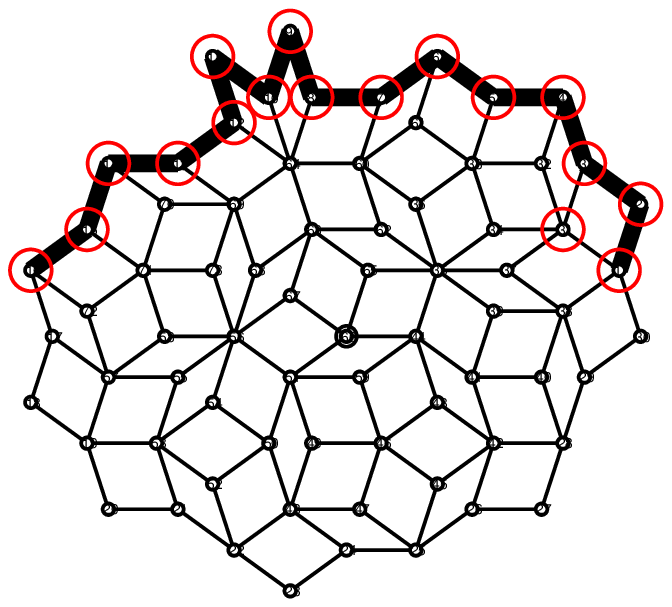,width=0.2\textwidth}\psfig{figure=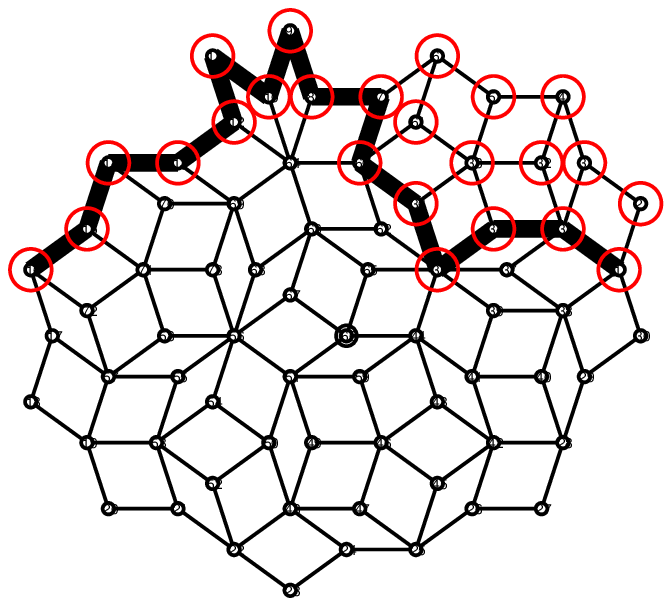,width=0.2\textwidth}
                  \psfig{figure=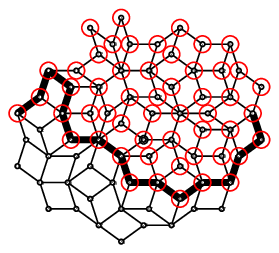,width=0.2\textwidth}\psfig{figure=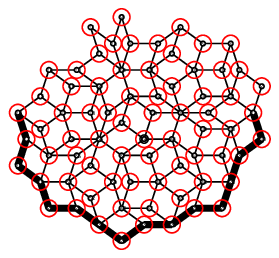,width=0.2\textwidth}}
      \caption{Shifting of a boundary line through a graph, from the initial (leftmost picture) to the final 
              (rightmost picture), corresponding to calculating the partition function
               by transfer-matrix method.\label{fig:boundarylines}}
      \end{figure}

\begin{figure}
      \centerline{\psfig{figure=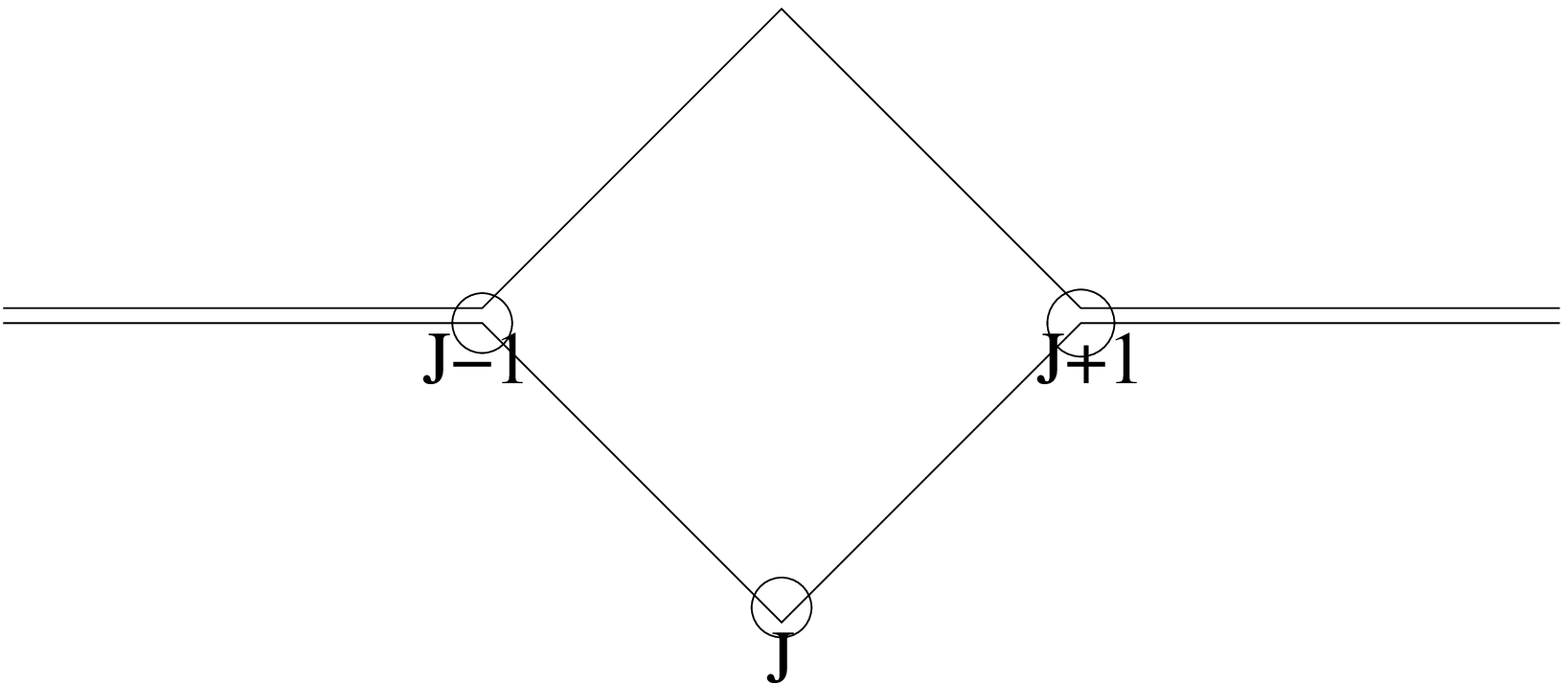,width=0.4\textwidth}
                  \psfig{figure=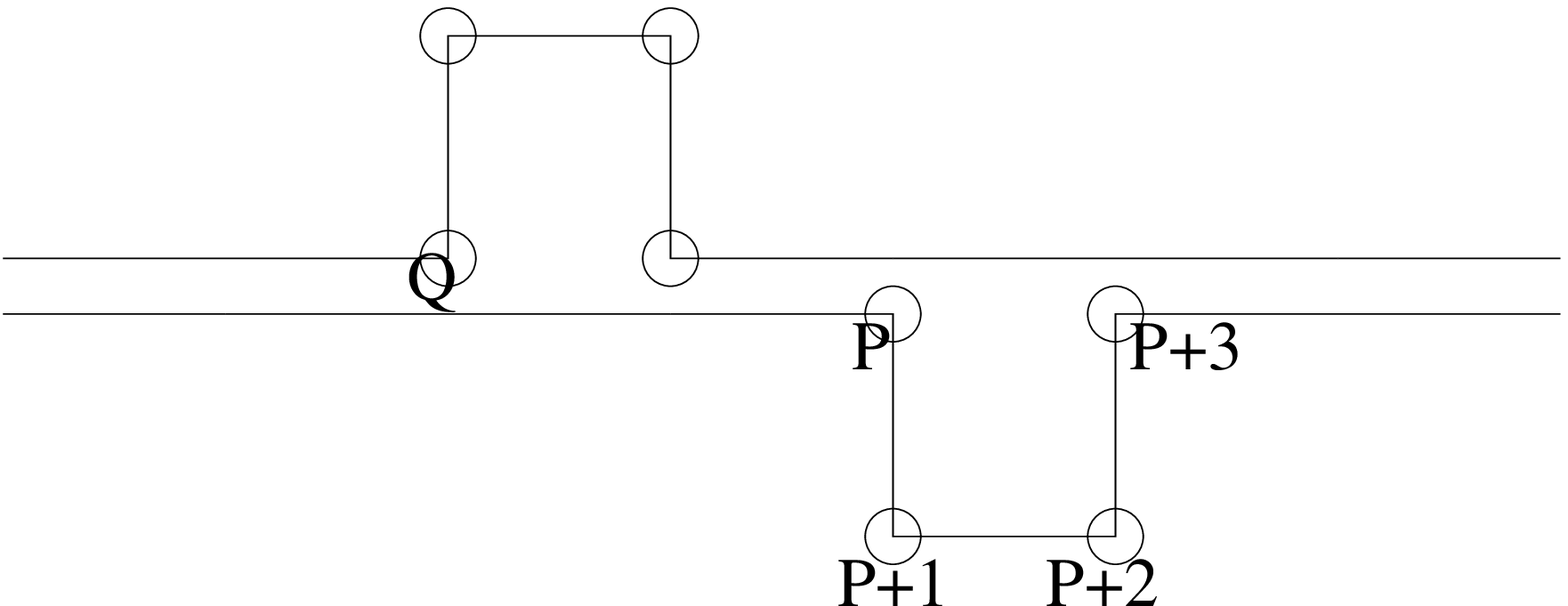,width=0.4\textwidth}
                 }
      \vspace{0.5cm}
      \caption{Two kinds of movements of the boundary line, 
               by one tile (left) and by two tiles (right).\label{fig:movements}}
\end{figure}

      Shifting the boundary line corresponds to 
      generating a new vector $Z'({\bf \sigma'})$ of 
      partition functions from the old vector $Z$. There is a lot of ambiguity in shifting the boundary line
      by a given number of tiles. In our case it amounts, however, to considering only three kinds of movements, 
      by one tile,
      by two tiles and a shift between two given boundary line configurations, which we discuss in the following. 
      Placing the initial boundary line on the perimeter
      of the graph and moving it at each stage by certain number of tiles, 
      see FIG.\ref{fig:boundarylines}, we have performed the sum over all configurations after reaching the final
      position of the boundary line (also lying on the perimeter).
      Now we discuss the details of updating the partition functions for the two kinds of boundary line movements,
      see FIG.\ref{fig:movements}. 
\subsubsection{One-tile movement}  For $1 \le J \le N$ we have
\begin{equation}
Z'({\bf \sigma}') \;=\; \tilde{x}^a (1-\tilde{y})^b \sum_{\sigma_J=1}^q Z({\bf \sigma})
\label{eq:onetilemovement}
\end{equation}
where $a = \Delta(\sigma'_{J-1},\sigma'_J) + \Delta(\sigma'_J,\sigma'_{J+1})$
and
$b = f \Delta(\sigma'_J,0)$, $f=2$.
\subsubsection{Two-tiles movement}  For $1 \le L < P \le N$ we have
\begin{equation}
Z'({\bf \sigma}'_\rho) \;=\; \tilde{x}^a (1-\tilde{y})^b \sum_{\sigma_{P+1}=1}^q \sum_{\sigma_{P+2}=1}^q Z({\bf \sigma})
\label{eq:twotilemovement}
\end{equation}
where $a = \Delta(\sigma'_L,\sigma'_{P+1})+\Delta(\sigma'_{P+1},\sigma'_{P+2})+
          \Delta(\sigma'_{P+2},\sigma'_{L+1})+\Delta(\sigma'_{P},\sigma'_{P+3})$
and
$b = f_1 \Delta(\sigma'_{P+1},0) + f_2 \Delta(\sigma'_{P+2},0)$
where $f_1=f_2=2$
and the new spin configuration is permuted with respect to the old one
${\bf \sigma}'_\rho = \{\sigma'_{\rho_1},\sigma'_{\rho_2},\ldots,\sigma'_{\rho_N}\}$ and
\begin{equation}
\rho_p \;=\; \left\{ \begin{array}{rrrr}
                        p   &  p &\le& L \\
                        P+1 &  p &=  & L+1 \\
                        P+2 &  p &=  & L+2 \\
                        p-2 &  p &\ge& L+3
                      \end{array}        \right.
\end{equation}

\subsubsection{Shifting the boundary line to the final position}
\begin{figure}
      \centerline{\psfig{figure=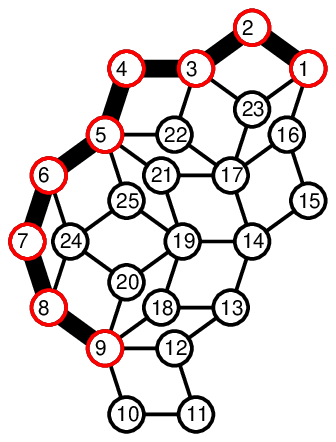,width=0.2\textwidth}
                  \psfig{figure=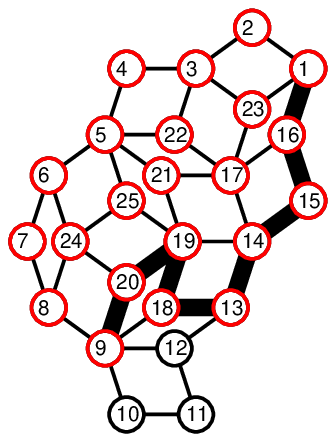,width=0.2\textwidth}
                  \psfig{figure=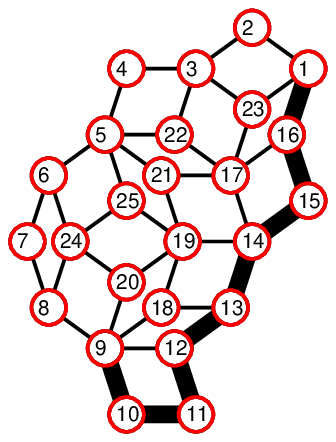,width=0.2\textwidth}
                 }
      \vspace{0.5cm}
      \caption{A boundary line (middle picture) which cannot be pushed forward by performing one of the movements
               discussed above. The initial and the final line configuration are shown at the left and at the right 
               respectively.\label{fig:dead_end}}
\end{figure}

\begin{figure}
\centerline{\psfig{figure=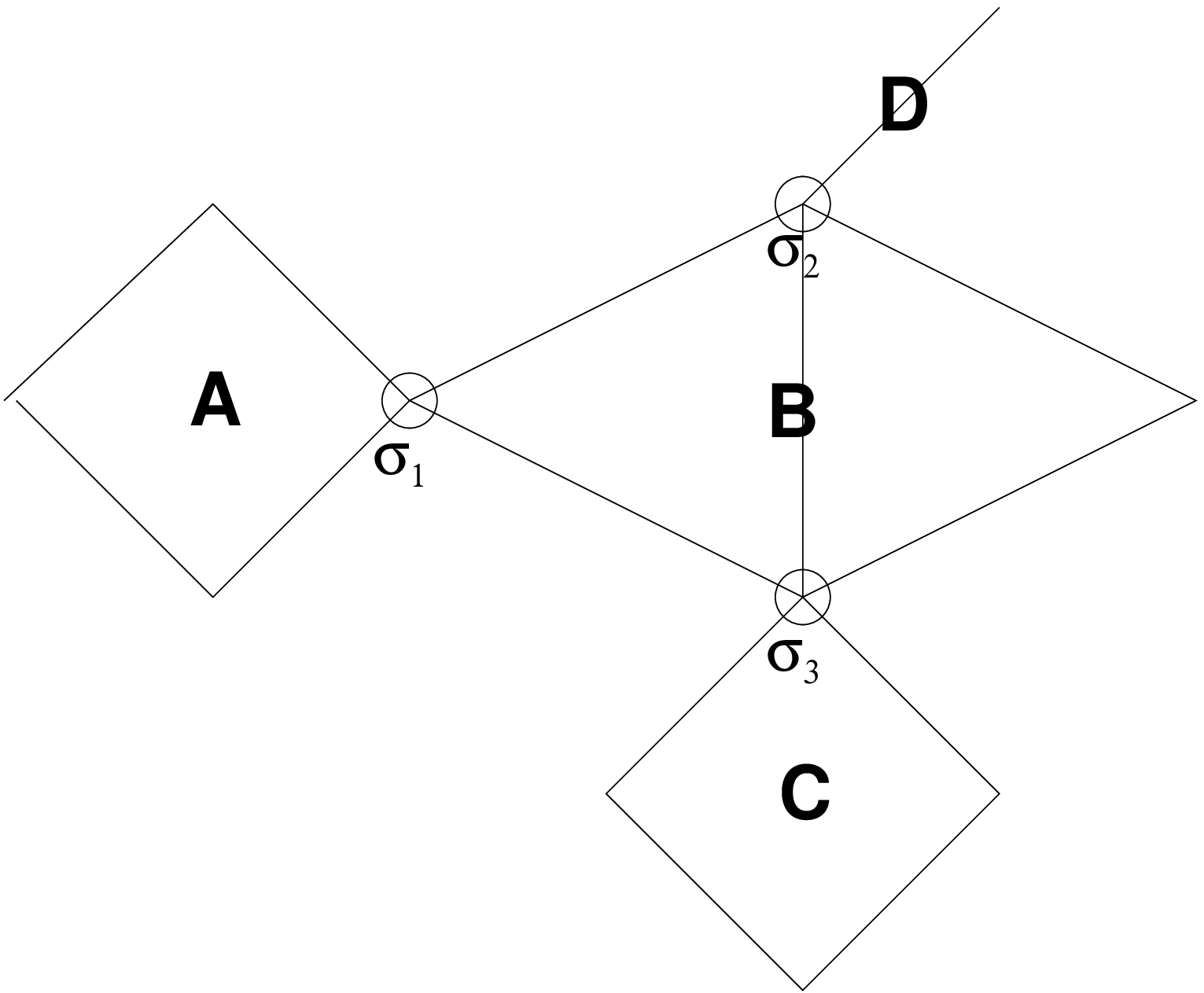,width=0.3\textwidth}
            \hspace{2cm}
            \psfig{figure=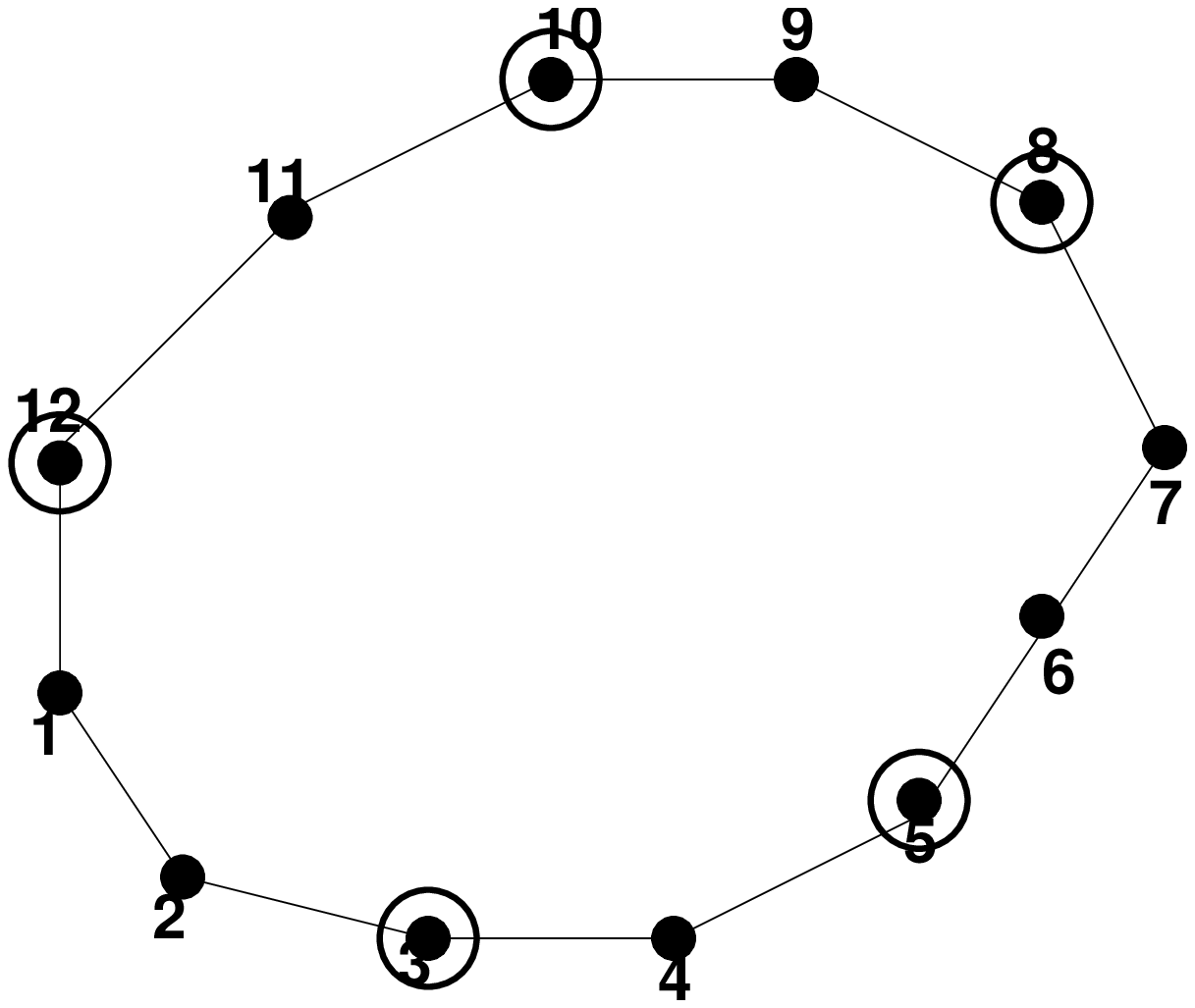,width=0.2\textwidth}}
\caption{A multicomponent graph consisting of four star components (left) 
         and the perimeter of a star graph with two isolated sites on the initial
         boundary line (consisting of sites from $1$ to $7$) and three isolated sites
         on the final boundary line (sites from $7$ to $12$) (right).
         The isolated sites are marked with circles.\label{fig:multicomponental_graph}}
\end{figure}

In most cases it is possible to displace the boundary line from the initial to the final position
by a sequence of the movements defined above. Sometimes, however, we arrive in a dead end because none 
of the movements can be done, see FIG.\ref{fig:dead_end}. In this case we have to shift the line 
directly to its final position by summing over all the spins which have not been taken into account yet.
The formal prescription for updating $Z({\bf \sigma})$ in this case reads:
\begin{equation}
Z'(\sigma_{F_1},\ldots,\sigma_{F_9}) \;=\; \sum_{\sigma_{s_1},\sigma_{s_2},\sigma_{s_3}} \tilde{x}^a (1-\tilde{y})^b
                        Z(\sigma_{B_1},\ldots,\sigma_{B_9})
\end{equation}
where $a = \sum_{j=1}^6 \Delta(\sigma_{e_{j,1}},\sigma_{e_{j,2}})$ and
      $b = 2 \sum_{j=1}^3 \Delta(\sigma_{s_{j}},0)$
and $B = \{1,16,15,14,13,18,19,20,9\}$, $F = \{1,16,15,14,13,12,11,10,9\}$, 
    $e = \{(9,10),(10,11),(11,12),(12,13),(12,9),(9,18)\}$ and $s = \{12,11,10\}$ denote
the current and the final boundary lines, the edges and the sites which were not taken into account yet respectively. 
Can the transfer-matrix formalism (tmf) also be applied to the case of a multicomponent graph?
The answer is affirmative because every connected graph can be dissected into its star graph components
for which the tmf is applicable. Since, however, star graph components share certain sites at their
boundaries, which we call in the following isolated sites, we have to calculate a whole set of partition functions
with given spin values at isolated sites and combine them to get the partition function of the whole graph.
In the following we assume the simplest case namely, that every isolated site is shared by exactly two star components. 
This was indeed the case by our overlap graphs.
Let us explain the procedure for the case of a graph depicted in FIG.\ref{fig:multicomponental_graph}.
The partition function $Z$ can be build up from partition functions 
$Z_A(\sigma_1)$, $Z_B(\sigma_1,\sigma_2,\sigma_3)$, $Z_C(\sigma_3)$ and $Z_D(\sigma_2)$ corresponding to star components
$A$, $B$, $C$, and $D$ with isolated spins $\sigma_1$, $\sigma_2$ and $\sigma_3$. 
\begin{equation}
Z \;=\; \sum_{\sigma_1,\sigma_2,\sigma_3} Z_A(\sigma_1) Z_B(\sigma_1,\sigma_2,\sigma_3) Z_C(\sigma_3) Z_D(\sigma_2)
\end{equation}
Now, the problem consists in calculating partition functions for a star graph
with specified spins at isolated sites located at the boundary. Assume that we have $p$ isolated
sites $j_k$, $k=1,..,p$ located at the initial boundary line and $q$ isolated sites
$l_k$, $k=1,..,q$ at the final boundary line respectively, see FIG.\ref{fig:multicomponental_graph}.
The calculation of $Z(s_{j_1},\ldots,s_{j_p},s_{l_1},\ldots,s_{l_q})$
amounts to repeating the tmf $Q^{p+q}$ times and modifying the initial and the final partition function set
by setting certain entries to zero. We replace the initial partition function set by
\begin{equation}
\left[ \prod_{k=1}^p \delta(\sigma_{j_k},s_{j_k}) \right] Z(\sigma_1,\ldots,\sigma_N)
\quad\mbox{for given}\quad s_{j_1},\ldots,s_{j_p}
\end{equation}
and the final partition function set is multiplied by 
$\left[ \prod_{k=1}^q \delta(\sigma_{l_k},s_{l_k}) \right]$ again for a given spin configuration
$s_{l_1},\ldots,s_{l_q}$. Another slight modification which is required consists in setting
the factors $f$,$f_1$ and $f_2$ entering in the exponent $b$ in equations 
(\ref{eq:onetilemovement},\ref{eq:twotilemovement}) according to whether the site is isolated (one)
or not (two).

\section{Series expansion of the free energy, magnetisation and field derivatives of the magnetisation}
\begin{figure}[!h]
\centerline{\psfig{figure=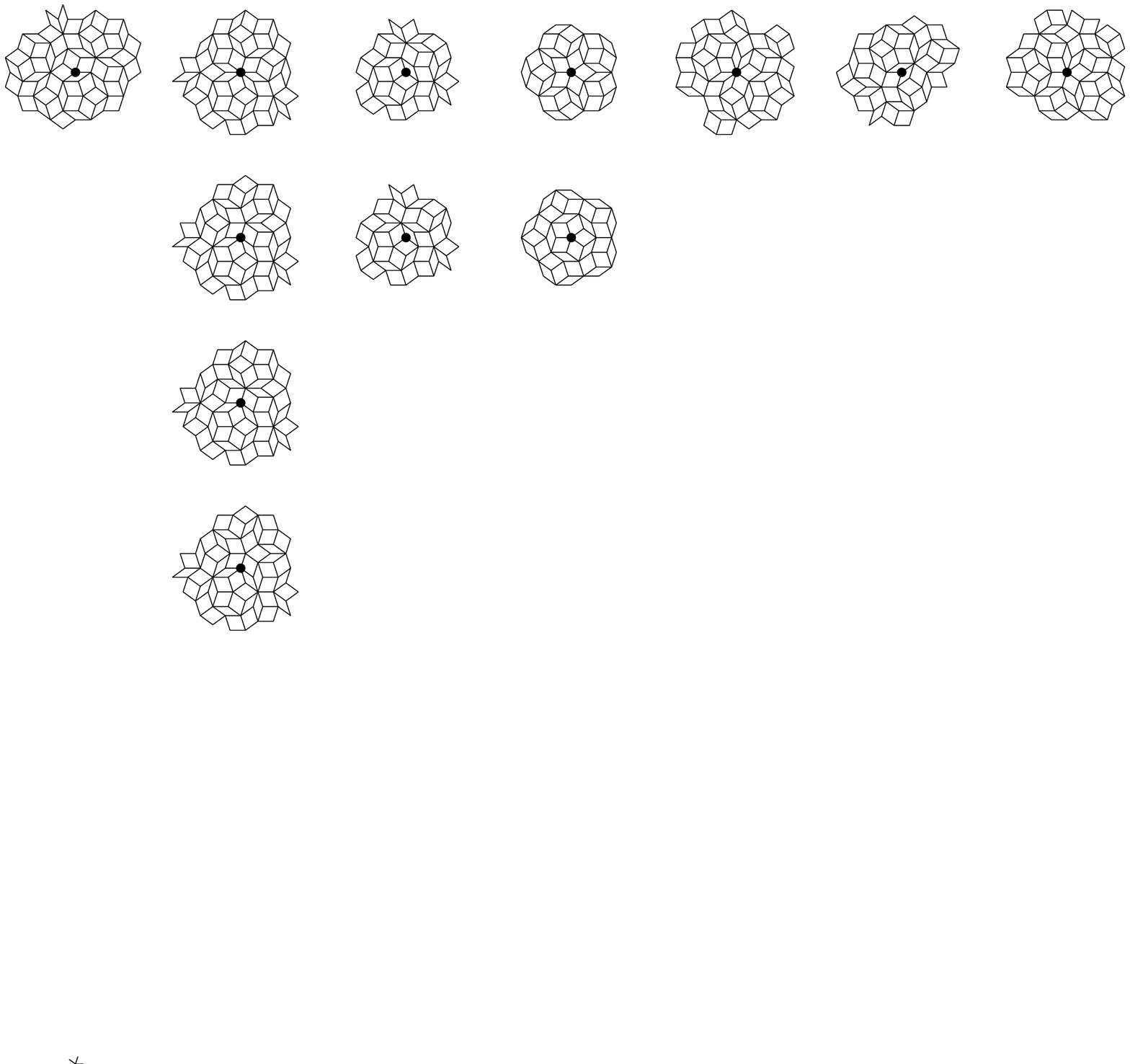,width=0.99\textwidth}}
\caption{Patches from the Penrose lattice used as input for the finite lattice method calculations.
         The columns contain all possible ``boundary line fillings'' of seven patches
         the central sites of which correspond to seven different vertex types.\label{fig:initial_patches}}
\end{figure}

We have performed calculations for a set of graphs constructed in the following way.
We cut off seven fairly round shaped patches from the PT 
so that the central sites of the  patches
had different vertex types and their perimeter lengths 
were not larger than $30$ edge lengths.
Then we enlarged the set of patches by all possible ``boundary line fillings'' obtaining in effect twelve
patches, see FIG.\ref{fig:initial_patches}.
In the next step we constructed all possible graphs contributing to the expansion in the recursive way described 
in section \ref{sec:Calculation}. Their number turned out to be $1004$.
This part of computations was rather tedious, up to two weeks for the second set on a SunOS machine,
because in generating graph overlaps many graphs turned up repeatedly and had to be rejected.
In the next step we generated ``decorations'' of graphs, i.e. we determined vertex types of all sites
of the graph including those on its boundary. Since the graphs could have several decorations 
the number of graphs we have to deal with increased to $5737$.
Now we were ready to compute the coefficients $a_r$ entering in (\ref{eq:high_temp_expans1})
which appeared to be different from zero only for a small fraction of all graphs, namely for 
$154$ graphs. This is not a surprising result since on the square lattice
the vast majority of rectangles used in the expansion yields zero coefficients as well \cite{Enting}.
Fortunately, most of the relevant graphs here were star graphs so we could easily compute the free energies
$\log{Z(g_r)}$ entering in (\ref{eq:high_temp_expans1}) in the way described in section \ref{sec:Calculation}.
There were however some awkward multicomponent graphs for which partition function computations
were more tedious. The series expansion is shown beneath.

After reordering the expansion (\ref{eq:high_temp_expans1}), i.e. collecting together terms
with the same power of $\tilde{y}$, the free energy $F(\tilde{x},\tilde{y})$ takes the form.
\begin{equation}
F(\tilde{x},\tilde{y}) \;=\; \log{Z({\cal G})} \;=\; F_0(\tilde{x}) + F_1(\tilde{x}) \tilde{y} + F_2(\tilde{x}) \tilde{y}^2 + \ldots \;=\; \sum_{n=0}^\infty F_n(\tilde{x}) \tilde{y}^n
\label{eq:high_temp_expans2}
\end{equation}
Quantities like the spontaneous magnetisation $M(\tilde{x})$, susceptibility $\chi(\tilde{x})$
and field derivatives of the susceptibility $\chi^{(n)}(\tilde{x}) = d^n \chi(\tilde{x})/d \tilde{y}^n$ can be expressed
as linear combinations of the polynomials $F_n(\tilde{x})$.
\begin{eqnarray}
M(\tilde{x}) \;=\; & \left. d F(\tilde{x},\tilde{y})/d B \right|_{B=0} & = F_1(\tilde{x}) \nonumber \\
\chi(\tilde{x}) \;=\; & \left. d^2 F(\tilde{x},\tilde{y})/d B^2 \right|_{B=0} & = 2 F_2(\tilde{x}) - F_1(\tilde{x}) \nonumber \\ 
\chi^{(1)}(\tilde{x}) \;=\; & \left. d^3 F(\tilde{x},\tilde{y})/d B^3 \right|_{B=0} & = 6 F_3(\tilde{x}) - 6 F_2(\tilde{x}) + F_1(\tilde{x}) \nonumber\\
\chi^{(2)}(\tilde{x}) \;=\; & \left. d^4 F(\tilde{x},\tilde{y})/d B^4 \right|_{B=0} & = 24 F_4(\tilde{x}) - 36 F_3(\tilde{x}) + 14 F_2(\tilde{x}) - F_1(\tilde{x})\nonumber\\
\end{eqnarray}
\begin{table}[!h]
\begin{tabular}{rrrrrrrrrrr}
       & $g_3$ & $g_4$ & $g_5$ & $g_6$ & $g_7$ & $g_8$ & $g_9$ & $g_{10}$ & $g_{11}$ \\
$M(x)$ & -1.0832 & -0.1803 & -1.3736 & -3.656 & -10.0258 & -12.8017 & -11.2619 & -16.135 & -72.9404\\
$\chi(x)$ & 2.1026 & 0.3607 & 4.4234 & 16.0886 & 63.0069 & 110.5216 & 137.4134 & 241.4194 & 990.5163\\
$\chi^{(1)}(x)$ & -4.0136 & -0.7214 & -15.448 & -73.4949 & -404.9629 & -941.0482 & -1484.2237 & -3081.2191 & -12969.5517\\
$\chi^{(2)}(x)$ & -33.5551 & -5.861 & -94.3 & -405.998 & -2039.6444 & -4479.9119 & -6807.027 & -13838.423 & -58117.8824\\
\end{tabular}
\begin{tabular}{rrrrrrrrrrr}
       & $g_{12}$ & $g_{13}$ & $g_{14}$ & $g_{15}$ & $g_{16}$ & $g_{17}$ & $g_{18}$ & $g_{19}$ & $g_{20}$ \\
$M(x)$ & -231.8001 & -493.5029 & -841.3831 & -1651.9956 & -4125.5155 & -9628.4929 & -18432.4073 & -33133.1842 & -70657.8814\\
$\chi(x)$ & 3397.510 & 8427.160 & 17816.643 & 41630.163 & 110568.346 & 277554.684 & 621098.121 & 1362853.02 & 3231583.007\\
$\chi^{(1)}(x)$ & -49115 & -139934 & -350428 & -944924 & -2738982 & -7510844 & -19047918 & -47924567 & -125178572\\
$\chi^{(2)}(x)$ & -217789 & -612298 & -1512009 & -4036132 & -11635964 & -31747510 & -79992423 & -200008381 & -520387269\\
\end{tabular}
\caption{Expansion coefficients of the magnetisation $M(x)$, susceptibility $\chi(x)$ 
         and its field derivatives $\chi^{(1)}(x)$, $\chi^{(2}(x)$ obtained from the finite lattice method. 
         \label{tab:magnetisation}}
\end{table}

\section{Verification of correctness of the computed expansion}
\begin{figure}
\centerline{\psfig{figure=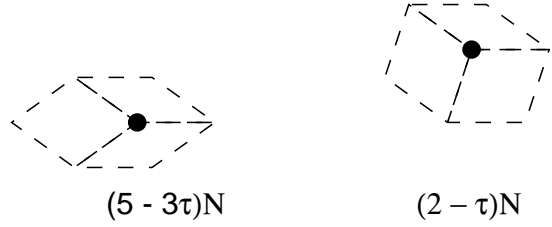,width=0.49\textwidth}}
\caption{Graphs from Penrose lattice contributing  to the coefficient $h_{3,1}$ 
         and their embedding numbers expressed through $\tau = (\sqrt{2} - 1)/2$.
         The dual graphs are constructed by connecting midpoints of rhombi
         abutting at bonds terminated by filled circles.\label{fig:coeff3_1}}
\end{figure}

\begin{figure}
\centerline{\psfig{figure=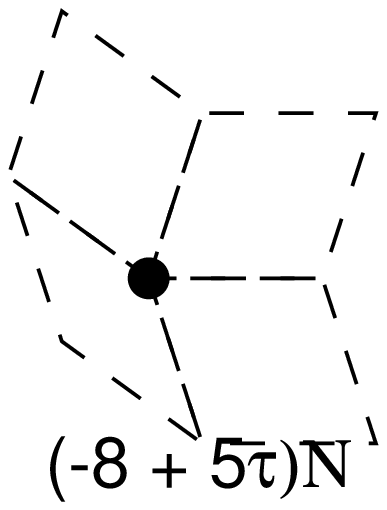,width=0.2\textwidth}
            \psfig{figure=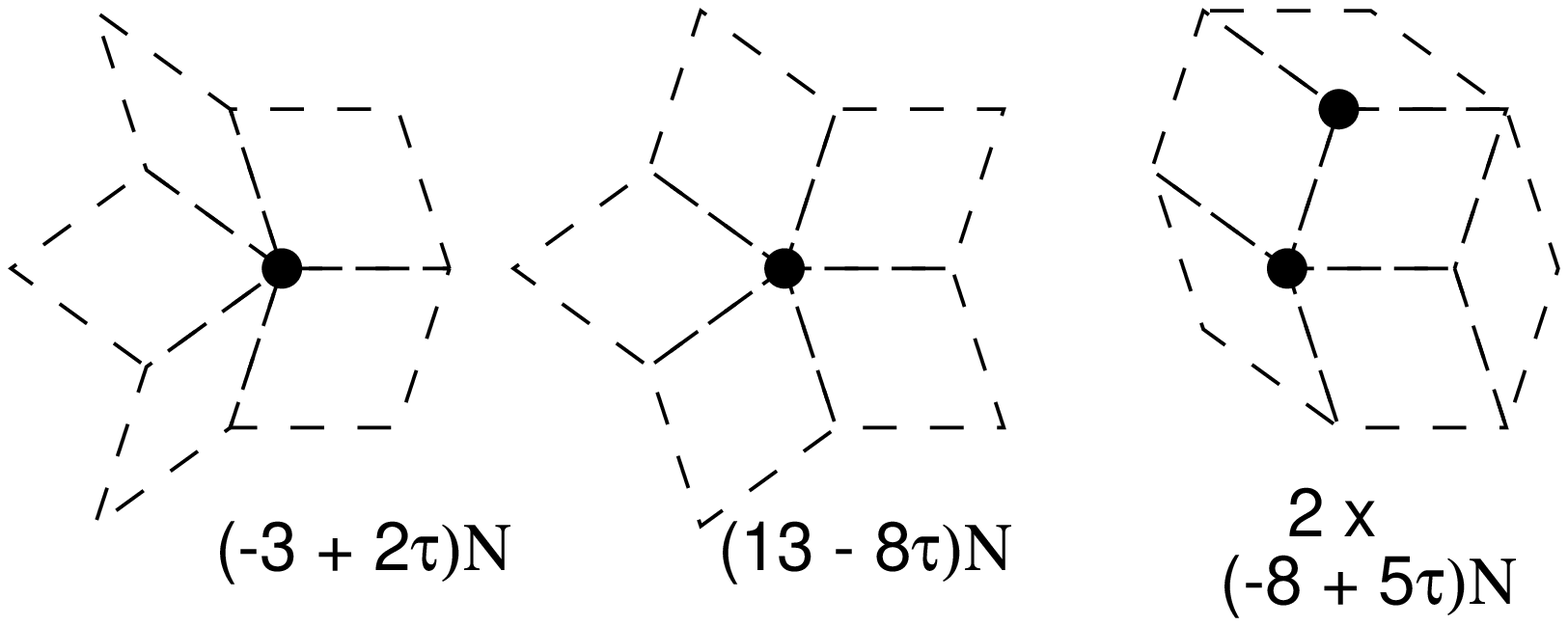,width=0.8\textwidth}}
\caption{The same as above corresponding  to coefficients $h_{4,1}$(three on the left) 
         and $h_{5,1}$(last on the right).\label{fig:coeff4_1}}
\end{figure}

\begin{figure}
\centerline{\psfig{figure=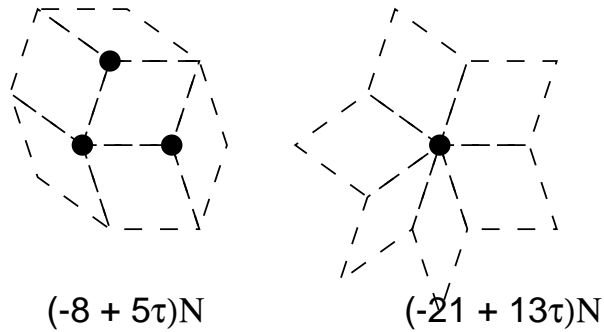,width=0.49\textwidth}}
\caption{Graphs contributing to the coefficients $h_{6,3}$(left) and $h_{6,1}$(right).\label{fig:coeff6_1}}
\end{figure}

\begin{figure}
\centerline{\psfig{figure=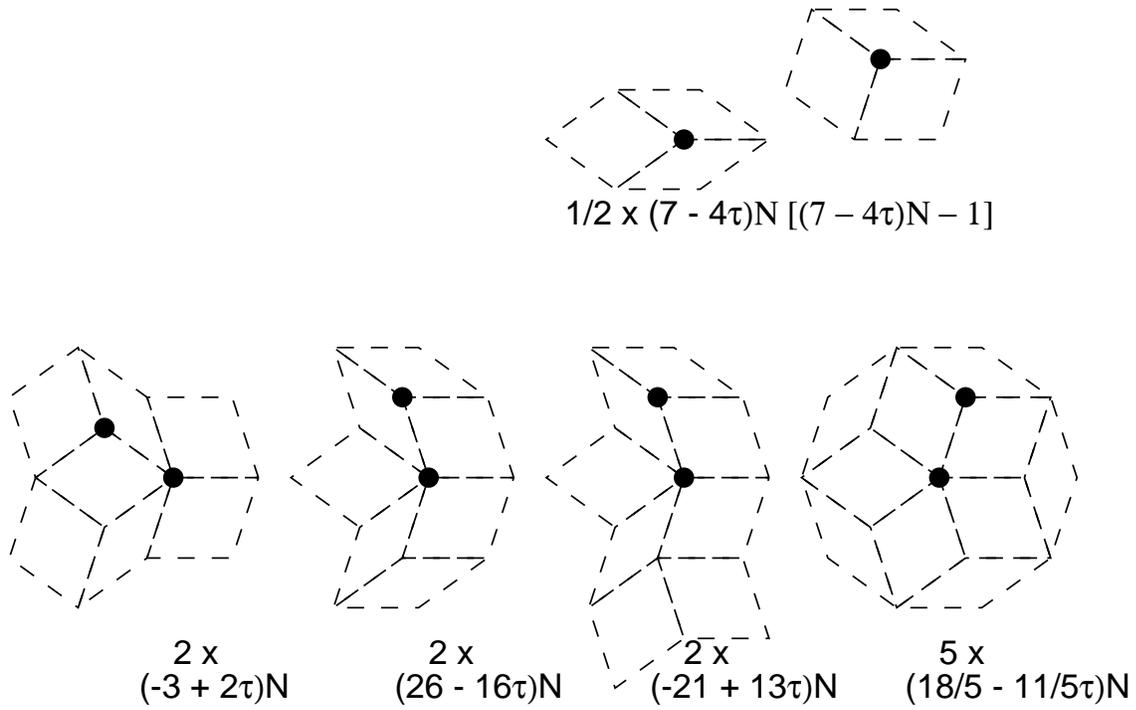,width=0.99\textwidth}}
\caption{Graphs contributing to the the coefficient $h_{6,2}$. \label{fig:coeff6_2}}
\end{figure}

There is a duality relation connecting the low temperature expansion of the Ising model on the
lattice ${\cal G}$ to the high-temperature expansion on the dual lattice ${\cal D}$, which 
takes the following form: 
\begin{equation}
Z_{\cal G}(x,y) \;=\; \exp\beta (M J + N B) \tilde{Z}_{\cal G}(x,y) 
                \;=\; 2^N (\cosh{\beta J})^M  (\cosh{\beta B})^N \tilde{Z}_{\cal D}(w,h) 
\label{eq:duality}
\end{equation}
where the low temperature variables are $x=\exp\{-2\beta J\}$, $y=\exp\{-\beta B\}$ and 
the high-temperature ones are $w=\tanh\{\beta J\}$, $h=\tanh\{\beta B\}$.
In the field free case $h=0$ the high-temperature expansion of $\tilde{Z}_{\cal D}(w,0)$ can be expressed 
by the square root of the determinant of a $2 M\times 2 M$ complex matrix \cite{RepetGrimm,Dolbilin}, 
which for periodic lattices
amounts to calculating a finite-dimensional determinant the dimension of which is of the
order of the size of the unit cell.    
Therefore the free energy expansion in variable $x$ can be calculated by taking
logarithms of equation (\ref{eq:duality}).
\begin{equation}
F \;=\; \lim_{N\longrightarrow \infty} \frac{1}{N} \log{Z_{\cal G}(x)}
  \;=\; \log{2} - \frac{q}{4}\log(1 - w^2) + \log{\tilde{Z}_{\cal D}(w)} 
\end{equation}
where $q = \lim_{N\longrightarrow \infty} 2 M/N$ is the mean coordination number.
The expansion of the last term on the right-hand side 
\newpage
\begin{equation}
\log{\tilde{Z}_{\cal D}(w)} \;=\; \sum_{n=3}^\infty g_n w^n
\label{eq:expansion_dual}
\end{equation}
is obtained from Kac-Ward determinants 
for large enough PAs of the Penrose lattice, see \cite{RepetGrimm} for detailed explanation.  
In TABLE \ref{tab:free_energy} we show the expansion coefficients $g_n$ for 
successive PAs together with the coefficients of $F_0(x)$ (see \ref{eq:high_temp_expans2}) obtained by the 
finite lattice method (FLM). 
The data for the highest approximants are quite close to these for the FLM;
the relative discrepancies for $n=3,\ldots,20$ are equal 
$-4.9\%$,$-0.2\%$,$-0.4\%$,$-5.1\%$,$-4.7\%$,$-2.8\%$,$6.8\%$,$14.5\%$,$1.\%$,$-1.6\%$,$-6.4\%$,$-9.3\%$,$-11.1\%$,$-9.4\%$,$-8.2\%$,
$-8.4\%$,$-11.3\%$,$-13.1\%$
and in most cases do not exceed ten percent. 
In addition both data sets depend on $n$ in a similar way.
Indeed, assuming known values for the critical point $x_c = 0.434269$ and the critical exponent $\alpha=2$ 
we define the sequence $r_n$ in the following way:
\begin{equation}
r_n \;=\; g_{n}/g_{n-1}  - 1/x_c(1 - (\alpha+1)/n)
\label{eq:residues}
\end{equation}
This sequence approaches zero $r_n \longrightarrow 0$ for large $n$, 
see pages 187--199 in \cite{DombGreenII}.
If we now compare the sequences from both the PA coefficients and
the FLM coefficients we see that the relative discrepancies
except for n=$5,10,12$ are all smaller than ten percent as well.
\begin{table}[!h]
\begin{tabular}{rrrrrrrrrrr}
m & 2 M  & $g_3$ & $g_4$ & $g_5$ & $g_6$ & $g_7$ & $g_8$ & $g_9$ & $g_{10}$ & $g_{11}$ \\
3 & 304 & 0.5132 & 0.1053 & 0.4868 & 0.8355 & 1.5658 & 1.3289 & 0.7368 & 1.0987 & 5.5921\\
4 & 796 & 0.5176 & 0.1106 & 0.4824 & 0.8116 & 1.5477 & 1.3518 & 0.7605 & 0.7286 & 4.8492\\
5 & 2084 & 0.5259 & 0.094 & 0.4722 & 0.8234 & 1.5873 & 1.358 & 0.6379 & 0.618 & 5.0768\\
6 & 5456 & 0.5279 & 0.0902 & 0.4707 & 0.8262 & 1.5894 & 1.3596 & 0.6452 & 0.6078 & 5.0249\\
7 & 14284 & 0.5279 & 0.0902 & 0.4716 & 0.8262 & 1.5858 & 1.3587 & 0.6551 & 0.6054 & 4.9812\\
8 & 37396 & 0.5277 & 0.0906 & 0.4672 & 0.8273 & 1.5467 & 1.3144 & 0.6159 & 0.5948 & 4.7844\\
9 & 97904 & 0.5266 & 0.09 & 0.4712 & 0.8207 & 1.5746 & 1.3443 & 0.6483 & 0.5732 & 4.8631\\ \hline
  &       & 0.5523 & 0.0902 & 0.473 & 0.8628 & 1.6486 & 1.3826 & 0.6043 & 0.4903 & 4.8161\\
\end{tabular}
\begin{tabular}{rrrrrrrrrr}
m & $g_{12}$ & $g_{13}$ & $g_{14}$ & $g_{15}$ & $g_{16}$ & $g_{17}$ & $g_{18}$ & $g_{19}$ & $g_{20}$ \\
3 & 14.3213 & 24.0789 & 33.3618 & 58.9491 & 134.8618 & 270.2105 & 413.0614 & 573.6842 & 1105.3993 \\
4 & 14.7944 & 26.4623 & 32.9548 & 50.7568 & 129.4246 & 287.3266 & 427.7404 & 495.4975 & 928.3412 \\
5 & 15.2147 & 26.4607 & 32.001 & 48.7386 & 128.2087 & 291.4299 & 432.9875 & 474.8081 & 872.2088 \\
6 & 15.1822 & 26.5257 & 31.9245 & 48.085 & 127.5097 & 292.4663 & 435.521 & 474.0257 & 866.5213 \\
7 & 15.1629 & 26.6334 & 32.006 & 47.7786 & 127.0104 & 292.7278 & 436.5489 & 473.8073 & 864.7229 \\
8 & 14.4285 & 25.1886 & 29.574 & 42.833 & 115.1099 & 269.0166 & 398.7356 & 415.9446 & 753.3294\\
9 & 14.8233 &         &        &        &          &          &          &          &         \\\hline
  & 15.0582 & 26.8024 & 32.3225 & 47.585 & 125.9445 & 290.9468 & 432.1051 & 463.0726 & 851.8084\\
\end{tabular}
\caption{Expansion coefficients for free energy (\ref{eq:expansion_dual}) for periodic approximants
         $m=3,\ldots,9$ of the dual Penrose lattice with $M$ edges in the unit cell. 
         Underneath the solid line coefficients obtained by the finite lattice expansion.
         The numbers $g_n$ approach
         the expansion coefficients for the dual Penrose lattice when $m\longrightarrow \infty$.\label{tab:free_energy}}
\end{table}
The lowest coefficients of our expansions can be also calculated exactly
by counting graphs on the dual Penrose lattice. Here we confine ourselves to the free energy and the magnetisation expansions.
We compute their first four nonzero coefficients and show that they are indeed close to those from tables 
TABLE \ref{tab:magnetisation} and TABLE \ref{tab:free_energy}.
We start from the low temperature expansion 
\begin{equation}
x^{M/2} y^{N} Z_{\cal G}(x,y) \;=\; \tilde{Z}_{\cal G}(x,y) \;=\; \sum_{n,m} h_{n,m} x^n y^{2m}
\end{equation}
with $h_{n,m}$ counting graphs, in general multicomponent graphs,
from dual lattice consisting of $m$ sites and $n$ bonds on the perimeter.
It is readily seen from figures (FIG.\ref{fig:coeff3_1}, FIG.\ref{fig:coeff4_1}, FIG.\ref{fig:coeff6_1} 
and FIG.\ref{fig:coeff6_2}) that
for the Penrose lattice the non zero coefficients take following values:
\begin{eqnarray}
h_{3,1} = (7 - 4\tau)    N & h_{4,1} = (-8 + 5\tau)   N  & \\ \nonumber
h_{5,1} = (10 - 6\tau)   N & h_{5,2} = (-16 + 10\tau) N  & \\ \nonumber
h_{6,1} = (-21 + 13\tau) N &  h_{6,3} = (-8 + 5\tau)   N   \\ \nonumber
\lefteqn{h_{6,2} = (22 -13\tau)N  + (7 - 4\tau) N ( (7 - 4\tau)N  - 1 )} \\ \nonumber
h_{7, 1} = (13 - 8\tau)   N & h_{7,4} = (-8 + 5\tau)   N &  \\ \nonumber
h_{7, 3} = (79 - 48\tau)  N & h_{7, 2} = (-63 + 39\tau) N + (7 - 4\tau)(-8 + 5\tau)N^2 & 
\label{eq:coeff}
\end{eqnarray}
\newpage
Let us notice that to coefficients $h_{6,2}$ and $h_{7, 2}$ contribute also disjoint, two-component
graphs thus the coefficients are second degree polynomials in $N$.

\begin{figure}
\centerline{\psfig{figure=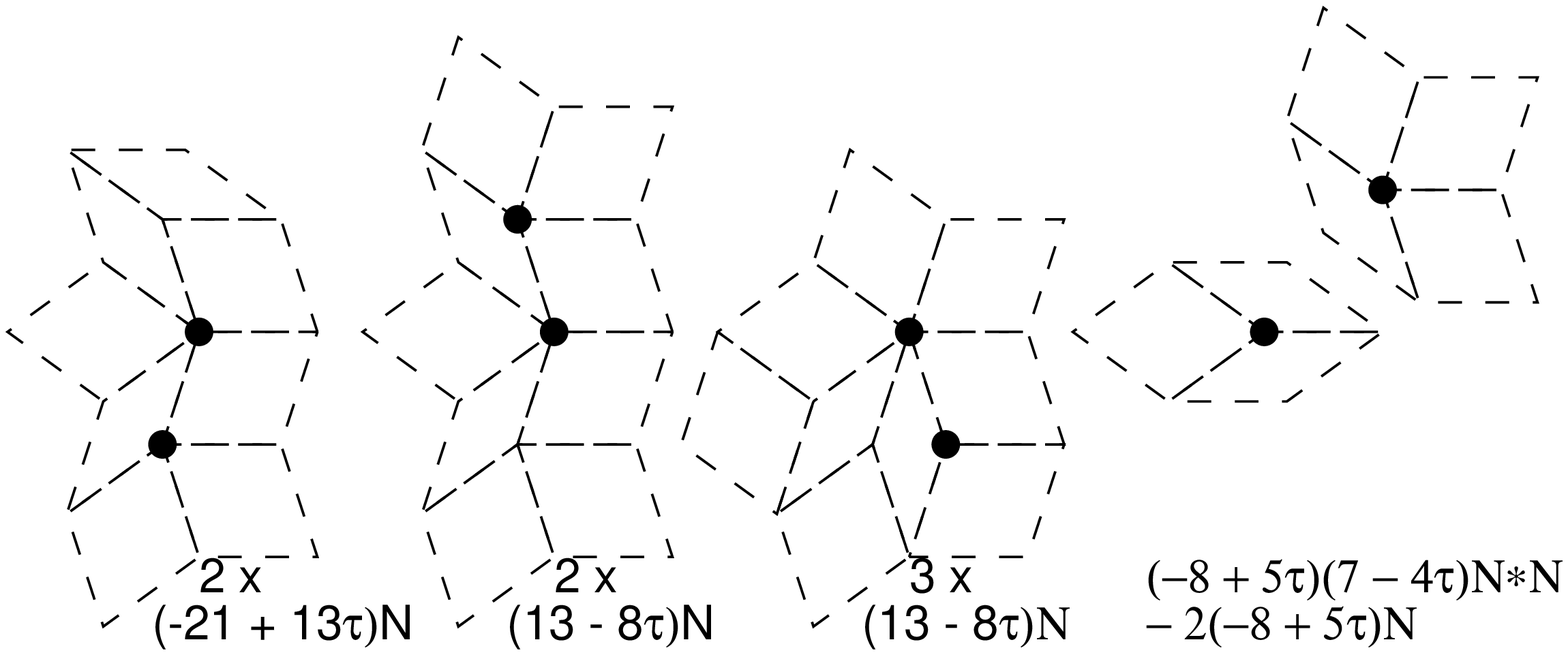,width=0.99\textwidth}}
\caption{Graphs contributing to the coefficient $h_{7,2}$. \label{fig:coeff7_2}}
\end{figure}
\begin{figure}
\centerline{\psfig{figure=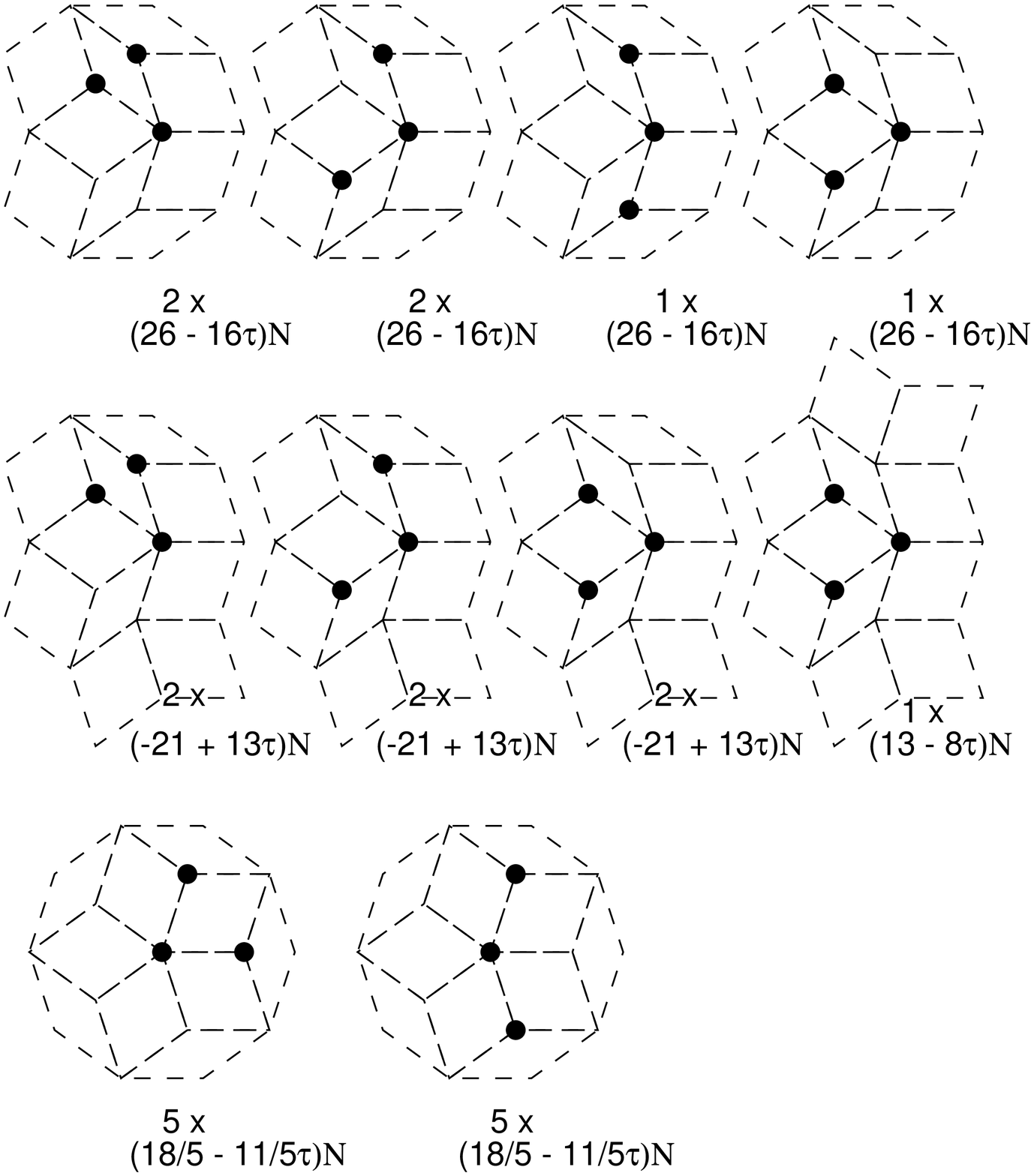,width=0.99\textwidth}}
\caption{Graphs contributing to the coefficient $h_{7,3}$. \label{fig:coeff7_3}}
\end{figure}
\begin{figure}
\centerline{\psfig{figure=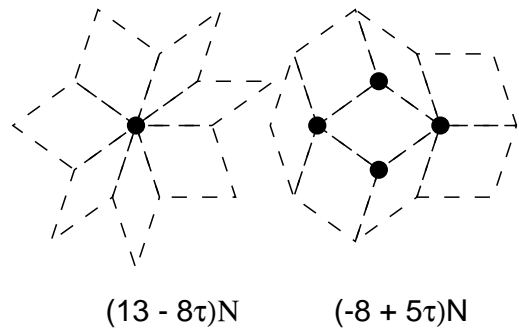,width=0.43\textwidth}}
\caption{Graphs contributing to the coefficients $h_{7,1}$(left) and $h_{7,4}$(right). \label{fig:coeff7_1}}
\end{figure}

If we now insert the coefficients from (\ref{eq:coeff}) into the definitions of 
the magnetisation $M(x)$ and the susceptibility $\chi(x)$
\begin{eqnarray}
M(x) &\;=\;& \left. \frac{1}{N} \frac{d \log[Z_{\cal G}(x,y)]}{d B} \right|_{B=0}  
     \;=\; \left. 1 - \frac{1}{N} y \frac{d \log[\tilde{Z}_{\cal G}(x,y)]}{d y} \right|_{y=1} \\ \nonumber
\chi(x) &\;=\;& \left. \frac{1}{N} \frac{d M(x)}{d B} \right|_{B=0} 
        \;=\; \left. \frac{1}{N}  y \frac{d}{d y} y \frac{d \log[\tilde{Z}_{\cal G}(x,y)]}{d y} \right|_{y=1}
\end{eqnarray}\\
we obtain following expansions:

\begin{tabular}{rrr}
$F(x)$ &=& $0.5279 x^3 + 0.0902 x^4 + 0.4721 x^5 + 0.8262 x^6 + 1.58359 x^7 + O(x^8)$ \\
$M(x)$ &=& $1 - 1.0557 x^3 - 0.1803 x^4 - 1.3049 x^5 - 3.4164 x^6 - 9.25233 x^7 + O(x^8)$ \\
$\chi(x)$ &=& $2.1115 x^3 + 0.3607 x^4 + 4.0526 x^5 + 14.6099 x^6 + 55.6843 x^7 + O(x^8)$ \\
$\chi^{(1)}(x)$ &=& $4.2229 x^3 + 0.7214 x^4 +  13.8761 x^5 + 64.6563 x^6 + 341.449 x^7 + O(x^8)$ \\
\end{tabular}\\
which conform quite well to the values from TABLE \ref{tab:magnetisation}.
Indeed, the relative differences between both sets of coefficients do not exceed ten percent in any case. 

\section{Asymptotic analysis of the series expansions}
\begin{figure}
      \centerline{\psfig{figure=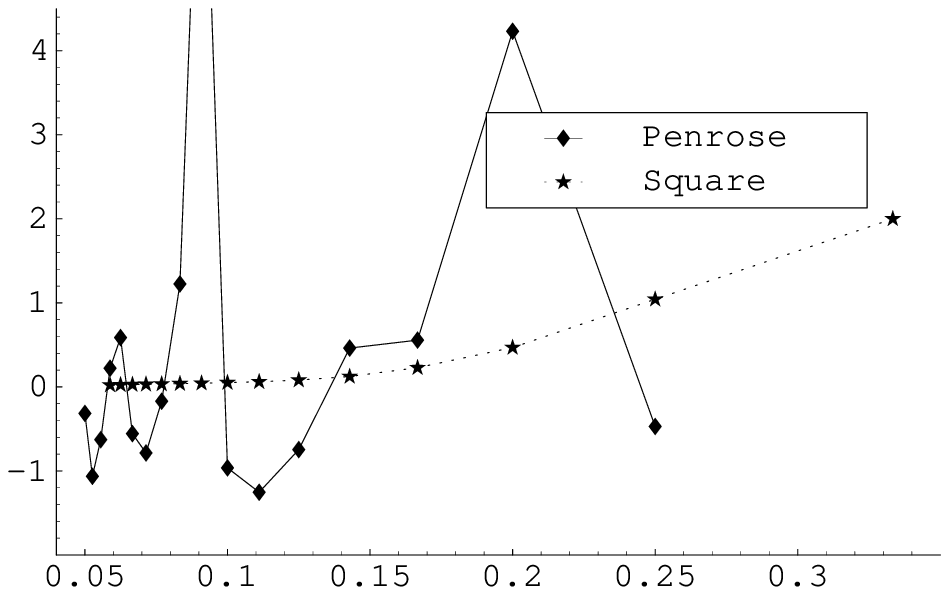,width=0.49\textwidth}
                  \psfig{figure=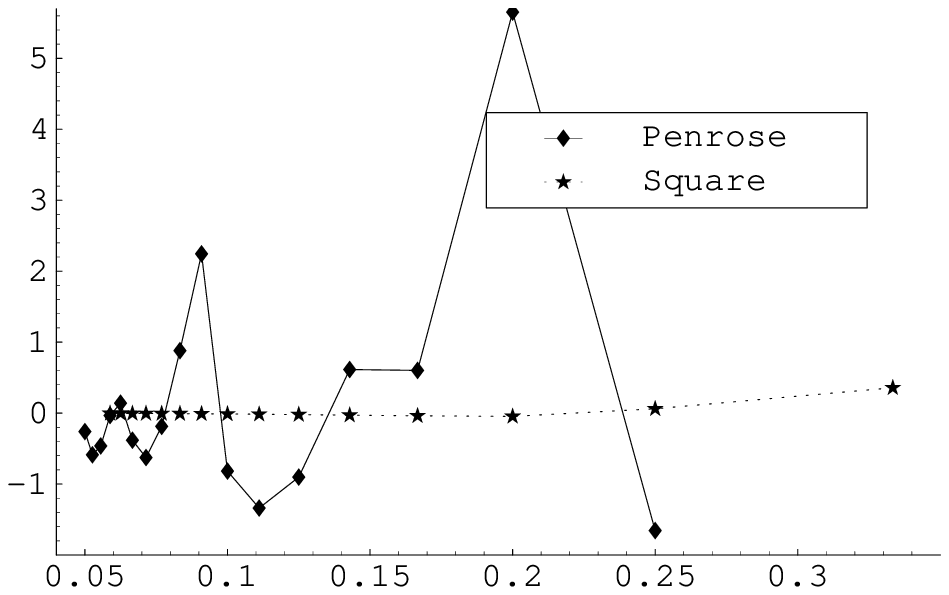,width=0.49\textwidth}}
      \caption{Plots of residues $r_n$ (\ref{eq:residues}) as a function of $1/n$
               for the free energy (left) and the magnetisation (right) on the Penrose and the square lattice respectively.
               In both cases we took the critical exponents $\alpha=2$ and $\beta=1/8$.\label{fig:QuotientPlots1}}
      \end{figure}

\begin{figure}
      \centerline{\psfig{figure=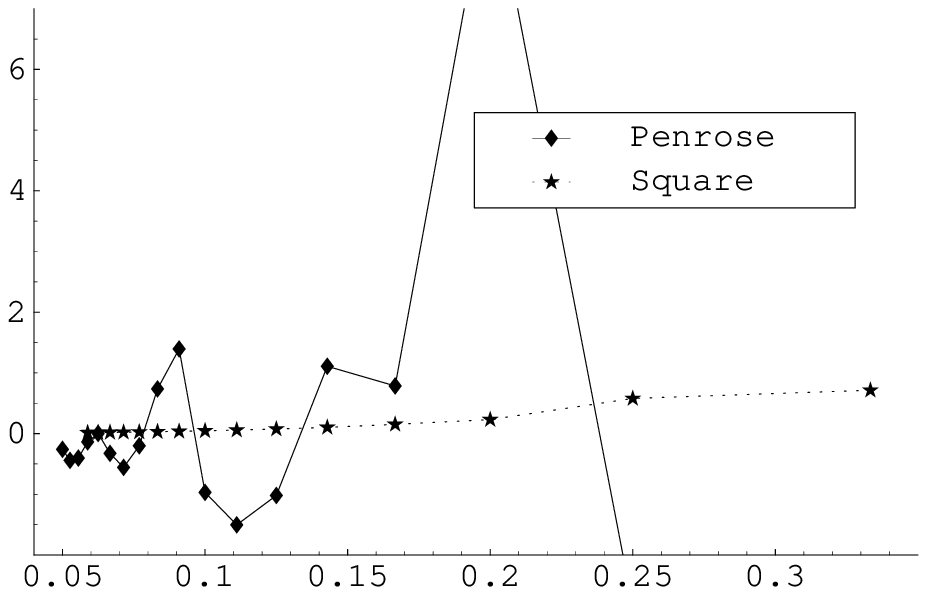,width=0.49\textwidth}
                  \psfig{figure=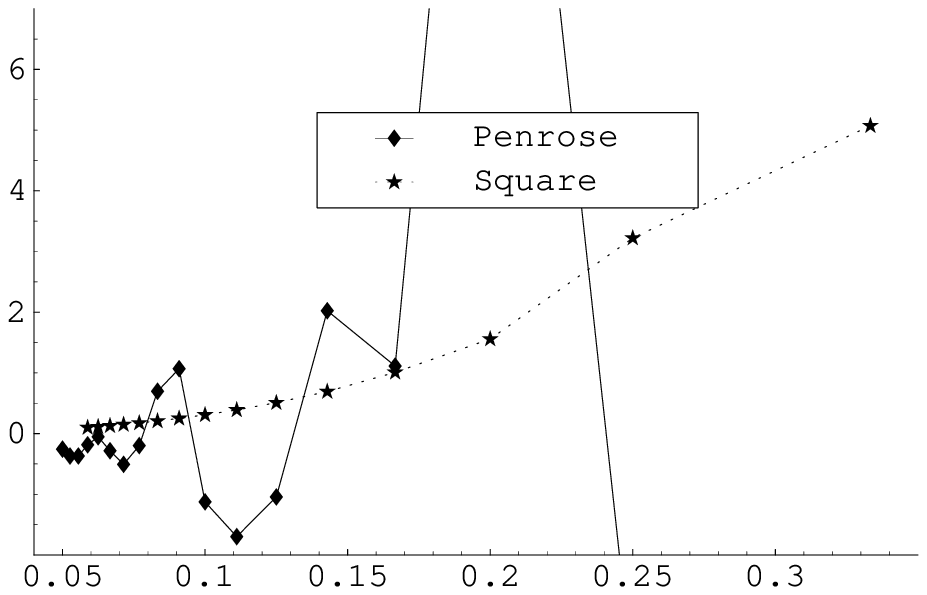,width=0.49\textwidth}}
      \caption{
               As before for the susceptibility (left) and the field derivative of susceptibility (right).
               In both cases we took the critical exponents $\gamma=-7/4$ and $\delta=-29/8$.\label{fig:QuotientPlots2}}
      \end{figure}

Now the problem consists in extracting critical exponents from the obtained expansions.
The simplest approach, the ratio method, in which one examines the asymptotically linear 
dependence of ratios $g_{n}/g_{n-1}$ 
(\ref{eq:residues}) on $1/n$  and obtains $x_c$ and $\alpha$ from linear regression,
is inapplicable in this case because of the slow convergence of series.
Indeed, the residues $r_n$ (\ref{eq:residues}) are much larger than those for the square
lattice and alternate in sign, see figures FIG.\ref{fig:QuotientPlots1} and FIG.\ref{fig:QuotientPlots2},
 what makes the asymptotic analysis difficult. This approach requires knowledge of $x_c$
which is known from other works \cite{Sorensen},\cite{PenLattFishZer} only with a limited accuracy.
Applying the Pad\'{e} method gives much more satisfactory results. 
Assuming that our thermodynamic functions $F(x)$ behave in the vicinity 
of the critical point $x_c$ like $F(x) \simeq (1 - x/x_c)^{-\alpha} A(x)$ it is readily seen that
functions  $G_0(x)$ and $G_1(x)$ behave asymptotically as follows
\begin{eqnarray}
G_0(x) & = & \frac{d}{d x}\left( \log \frac{d F(x)}{d x}\right) \left/ \right. \frac{d}{d x} \left( \log F(x)\right)  
     \simeq \frac{\alpha + 1}{\alpha} \;+\; O(x - x_c) \\ \nonumber
G_1(x) & = & (x - x_c) \frac{d}{d x} \left( \log F(x)\right) \simeq \alpha \;+\; O(x - x_c)
\label{eq:PadeApprox} \\
\end{eqnarray}
Constructing Pad\'{e} approximants to $G_0(x)$ and $G_1(x)$ and evaluating them at $x=x_c=0.434269$ usually yields a reasonable estimation 
of $\alpha$, even if $x_c$ is only known with a moderate accuracy. 
Results of this analysis, shown in tables TABLE \ref{tab:PadeExponI} and TABLE \ref{tab:PadeExponII}
are fairly close to the square lattice exponents. The sequence of Pad\'{e} approximants is quite
stable, except for few cases  marked by asterix.  Whenever the square lattice results converged Penrose data did as well. 

On the other hand we can compute biased estimates of $x_c$, assuming known values of critical exponents.
Indeed, the appropriate poles of Pad\'{e} approximants to  the function $[F(x)]^{1/\alpha}$ should give rapidly
convergent sequence of estimates of $x_c$. These sequences for the magnetisation expansion are shown in TABLE \ref{tab:PadeExponIII}.
In most cases the data do not deviate more than one percent 
from the exact values $x_c=0.434269$ (Penrose lattice \cite{RepetGrimm1}) and $x_c=\sqrt(2)-1$ (square lattice).

We can therefore claim that the data supports the claim that the quasiperiodic Ising model under consideration belongs to the square 
lattice (Onsager) universality class.
\begin{table}[!h]
\begin{tabular}{r|rr|rr|rr|rr|rr|rr} 
n &  \multicolumn{12}{c}{Approximant} \\
  &  \multicolumn{4}{c}{[n,n-1]} & \multicolumn{4}{c}{[n,n]} & \multicolumn{4}{c}{[n,n+1]} \\ \hline
  &  \multicolumn{2}{c}{Penrose} & \multicolumn{2}{c}{Square}&  \multicolumn{2}{c}{Penrose} & \multicolumn{2}{c}{Square}      &  \multicolumn{2}{c}{Penrose} & \multicolumn{2}{c}{Square}        \\ 
  & $G_0(x)$ & $G_1(x)$ & $G_0(x)$ & $G_1(x)$ & $G_0(x)$ & $G_1(x)$ & $G_0(x)$ & $G_1(x)$ & $G_0(x)$ & $G_1(x)$ & $G_0(x)$ & $G_1(x)$ \\ \hline
$8$ & $0.159^{*}$ & $0.128$ & $0.125$ & $0.125$ & $0.144$ & $0.126$ & $0.125$ & $0.125$ & $0.126$ & $0.126$ & $0.125$ & $0.125$ \\
$9$ & $0.242^{*}$ & $0.126$ & $0.125$ & $0.125$ & $0.125$ & $0.126$ & $0.125$ & $0.125$ & $0.126$ & $0.127$ & $0.125$ & $0.125$ \\
$10$ & $0.121$ & $0.127$ & $0.125$ & $0.125$ & $0.123$ & $0.126$ & $0.125$ & $0.125$ & $0.126$ & $0.126$ & $0.125$ & $0.125$ \\
$11$ & $0.121$ & $0.126$ & $0.125$ & $0.125$ & $0.125$ & $0.126$ & $0.125$ & $0.125$ & $0.126$ & $0.124$ & $0.125$ & $0.125$ \\
$12$ & $0.140$ & $0.128$ & $0.125$ & $0.125$ & $0.135$ & $0.114$ & $0.125$ & $0.125$ & $0.113$ & $0.120$ & $0.125$ & $0.125$ \\
$13$ & $0.139$ & $0.121$ & $0.125$ & $0.125$ & $0.115$ & $0.118$ & $0.125$ & $0.125$ & $0.113$ & $0.122$ & $0.125$ & $0.125$ \\
$14$ & $0.067^{*}$ & $0.128$ & $0.125$ & $0.125$ & $0.042^{*}$ & $0.097^{*}$ & $0.125$ & $0.125$ & $0.131$ & $0.137$ & $0.125$ & $0.125$ \\
$15$ & $0.069^{*}$ & $0.154^{*}$ & $0.125$ & $0.125$ & $0.128$ & $0.146$ & $0.125$ & $0.125$ & $0.130$ & $0.100$ & $0.125$ & $0.125$ \\
\end{tabular}
\caption{Estimates of the magnetisation critical exponent $\beta$  by means of Pad\'{e} approximants $[n,m]$
         to functions $G_0(x)$ and $G_1(x)$ (\ref{eq:PadeApprox}) constructed from the expansion to order $20$ 
         for the square- and Penrose lattice respectively. 
         Entries marked by asterix differ strongly from square lattice exponents.
         \label{tab:PadeExponI}}
\end{table}
\begin{table}[!h]
\begin{tabular}{r|rr|rr|rr|rr|rr|rr} 
n &  \multicolumn{12}{c}{Approximant} \\
  &  \multicolumn{4}{c}{[n,n-1]} & \multicolumn{4}{c}{[n,n]} & \multicolumn{4}{c}{[n,n+1]} \\ \hline
  &  \multicolumn{2}{c}{Penrose} & \multicolumn{2}{c}{Square}&  \multicolumn{2}{c}{Penrose} & \multicolumn{2}{c}{Square}      &  \multicolumn{2}{c}{Penrose} & \multicolumn{2}{c}{Square}        \\ 
  & $G_0(x)$ & $G_1(x)$ & $G_0(x)$ & $G_1(x)$ & $G_0(x)$ & $G_1(x)$ & $G_0(x)$ & $G_1(x)$ & $G_0(x)$ & $G_1(x)$ & $G_0(x)$ & $G_1(x)$ \\ \hline
$ 8$ & $-1.287^{*}$ & $-1.855    $ & $-1.383^{*}$ & $-1.743$ & $-1.469$ & $-1.778$ & $-1.741$ & $-1.740$ & $-1.095^{*}$ & $-2.230$ & $-1.741$ & $-1.741$ \\
$ 9$ & $-1.313^{*}$ & $-1.832    $ & $-1.741$ & $-1.741$ & $-1.397^{*}$ & $-1.946$ & $-1.741$ & $ 0.000^{*}$ & $-0.808^{*}$ & $-2.244$ & $-1.699$ & $-1.585$ \\
$10$ & $-1.299^{*}$ & $-1.834    $ & $-1.702$ & $-1.585$ & $-1.723$ & $-1.790$ & $-1.716$ & $-1.454$ & $-1.259$ & $-1.559$ & $-1.716$ & $-1.587$ \\
$11$ & $-1.431    $ & $-1.851    $ & $-1.716$ & $-1.587$ & $-1.133^{*}$ & $-1.249^{*}$ & $-1.716$ & $ 0.000^{*}$ & $-1.088^{*}$ & $-1.758$ & $-1.732$ & $-1.747$ \\
$12$ & $-1.095^{*}$ & $-3.743^{*}$ & $-1.738$ & $-1.747$ & $-1.131^{*}$ & $-0.244^{*}$ & $-1.736$ & $-1.747$ & $-1.259^{*}$ & $-1.129^{*}$ & $-1.736$ & $-1.747$ \\
$13$ & $-1.472    $ & $-1.569    $ & $-1.736$ & $-1.747$ & $-2.005$ & $-0.956^{*}$ & $-1.736$ & $ 0.000^{*}$ & $13.636^{*}$ & $ 0.316^{*}$ & $-1.757$ & $-1.748$ \\
$14$ & $-0.336^{*}$ & $-0.752^{*}$ & $-1.737$ & $-1.747$ & $-4.638^{*}$ & $-2.848^{*}$ & $-1.742$ & $-1.748$ & $-4.610^{*}$ & $-2.432^{*}$ & $-1.742$ & $-1.748$ \\
$15$ & $-4.610^{*}$ & $-2.458^{*}$ & $-1.742$ & $-1.748$ & $-4.637^{*}$ & $-1.924$ & $-1.742$ & $-0.001^{*}$ & $-7.190^{*}$ & $-2.411^{*}$ & $-1.743$ & $-1.749$ \\
\end{tabular}
\caption{As before for the susceptibility critical exponent $\gamma$. 
         \label{tab:PadeExponII}}
\end{table}
\begin{table}[!h]
\begin{tabular}{r|r|r|r|r|r|r} 
n &  \multicolumn{6}{c}{Approximant} \\
  &  \multicolumn{2}{c}{[n,n-1]} & \multicolumn{2}{c}{[n,n]} & \multicolumn{2}{c}{[n,n+1]} \\ \hline
  &  \multicolumn{1}{c}{Penrose} & \multicolumn{1}{c}{Square}&  \multicolumn{1}{c}{Penrose} & \multicolumn{1}{c}{Square}      &  \multicolumn{1}{c}{Penrose} & \multicolumn{1}{c}{Square}        \\ \hline
$8$ &  $0.434$ & $0.414$ & $0.462$ & $0.414$ & $0.437$ & $0.414$ \\
$9$ &  $0.433$ & $0.414$ & $0.434$ & $0.414$ & $0.434$ & $0.414$ \\
$10$ &  $0.434$ & $0.414$ & $0.434$ & $0.414$ & $0.434$ & $0.414$ \\
$11$ &  $0.434$ & $0.414$ & $0.434$ & $0.414$ & $0.435$ & $0.414$ \\
$12$ &  $0.433$ & $0.414$ & $0.446$ & $0.414$ & $0.437$ & $0.414$ \\
$13$ &  $0.434$ & $0.414$ & $0.435$ & $0.414$ & $0.436$ & $0.414$ \\
$14$ &  $0.431$ & $0.414$ & $0.438$ & $0.414$ & $0.457 + 0.018*I$ & $0.414$ \\ \hline
\end{tabular}
\caption{Biased estimates of the critical point $x_c = \exp(-2\beta_c)$ obtained from the magnetisation
         expansion to order $20$  for the square- and Penrose lattice respectively.\label{tab:PadeExponIII}}
\end{table}

\section{Concluding remarks and outlook}
\begin{figure}
      \centerline{\psfig{figure=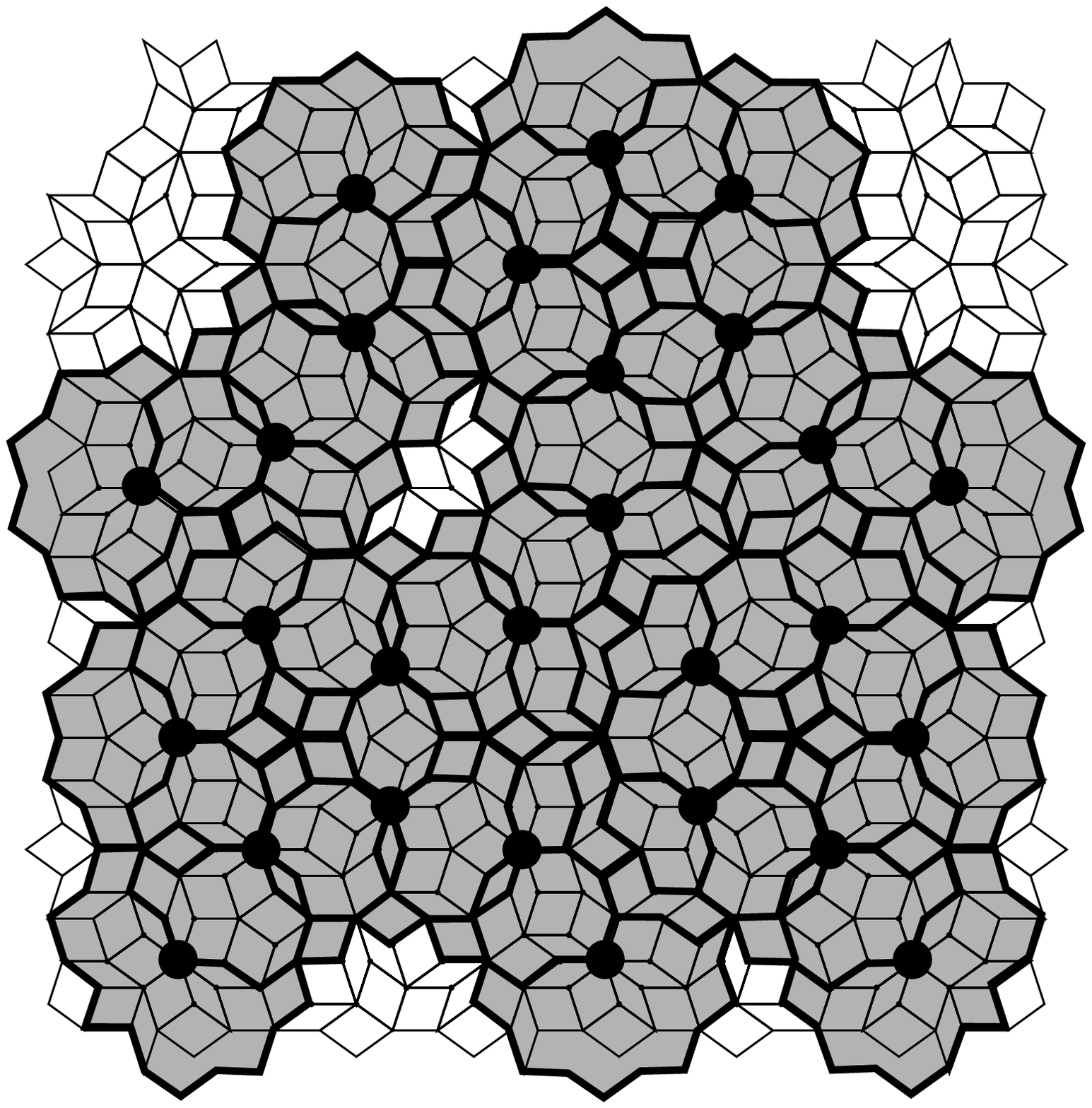,width=0.75\textwidth}
                  \psfig{figure=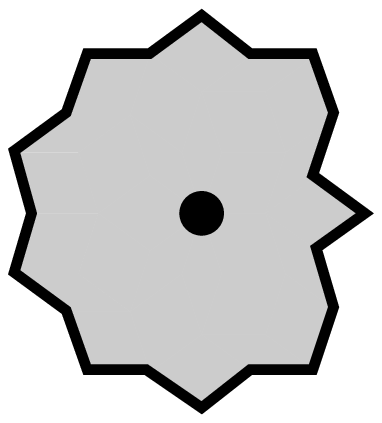,width=0.25\textwidth}}
      \caption{Covering of a fragment of the Penrose lattice (left) with copies of a patch
               consisting of $18$ thick and $10$ thin rhombi (right). Central sites of patches
               are marked by black dots. One can see that the covering is not complete, there are holes
               not covered by any of the patches.\label{fig:covering}}
      \end{figure}

The aim of this work was to analyse quasiperiodic Ising models by means of graphical series
expansions. We calculated low temperature expansions of the free energy, magnetisation and
the susceptibility to order $20$ and extracted the respective critical exponents.
We note that we did not obtain exact values for the exponents.
They are marked by errors, which however do not exceed 10\% in most cases. 
This feature is rather unusual for series expansions on regular lattices
where the coefficients are exact to a given order which is determined by the size of patches 
used for calculations. This deserves further comment.
There are the following two sources of errors in our method:
\begin{enumerate}
\item There are always graphs which cannot be embedded in any of the patches used in the calculations.
\item The FLM consists in covering and probing the lattice with a finite set
      of patches. Since the lattice is irregular, in particular, not periodic there can always
      be ``holes'', i.e. groups of sites, associated with graphs on the dual lattice,
      which are not covered by any of the patches. Such incomplete covering with only one single
      patch is shown in figure \ref{fig:covering}. Consequently, even if the patches are large
      there can always be few small graphs which cannot be embedded into them and which 
      produce an error in even the lowest order coefficients.
\end{enumerate}
\newpage

The next step in our research is to analyse the quasiperiodic $Q$-state Potts models,
especially for $Q=3,4$ because the Harris-Luck criterion implies that quasiperiodic order
should be strong enough to alter the critical behaviour in these cases.
The main problem here consists in calculating partition functions for finite patches.
This can be done, for instance, by improving the FLM \cite{ImprovedFLM} where the expansion
for a particular $Q$ is obtained from partition functions with smaller $Q$ values.
Another possibility is to use the Fortuin-Kasteleyn representation of the Potts model \cite{PottsModel}, 
expressing partition functions as a function of $Q$.
Work in this direction is in progress.

\section{Acknowledgements}
 P.R. thanks Des Johnston for discussions and an attentive proof reading of the paper.
 This work has been supported by a European Community IHP network HPRN-CT-1999-00161 ``EUROGRID''.

\end{document}